\documentclass[12pt]{article}

\usepackage{cite}

\usepackage{textcomp}
\usepackage[centertags]{amsmath}
\usepackage{amssymb}

\usepackage[bf]{caption}
\usepackage{graphicx}

\title{Structure formation in active networks}

\author
{Simone K\"{o}hler, Volker Schaller and Andreas R. Bausch$^\ast$\\
\\
\normalsize{Lehrstuhl f\"{u}r Biophysik E27, Technische Universit\"{a}t M\"{u}nchen,} \\
\normalsize{James-Franck-Stra\ss e 1, 85748 Garching, Germany}\\
\\
\normalsize{$^\ast$To whom correspondence should be addressed; E-mail:  abausch@ph.tum.de.}
}

\date{}

\begin{document}

\baselineskip24pt

\maketitle

\textbf{
Structure formation and constant reorganization of the actin cytoskeleton are key requirements for the function of living cells. Here we show that a minimal reconstituted system consisting of actin filaments, crosslinking molecules and molecular-motor filaments exhibits a generic mechanism of structure formation, characterized by a broad distribution of cluster sizes. We demonstrate that the growth of the structures depends on the intricate balance between crosslinker-induced stabilization and simultaneous destabilization by molecular motors, a mechanism analogous to nucleation and growth in passive systems. We also show that the intricate interplay between force generation, coarsening and connectivity is responsible for the highly dynamic process of structure formation in this heterogeneous active gel, and that these competing mechanisms result in anomalous transport, reminiscent of intracellular dynamics.}

\paragraph*{Introduction}
Cell adhesion, cytokinesis or cell migration are some of the fascinating examples of the power of the cytoskeleton to self organize locally into complex structures. While such processes are tightly controlled by the biochemical activation of the participating proteins, structures need to be formed by a physical self organization process.
A stable structure formation mechanism requires the intricate interplay between locally produced forces and the stabilization of the filamentous network by crosslinking molecules. Despite the fundamental importance of self organization processes in such active systems, the microscopic mechanisms and their consequences are poorly understood. To this end, the concept of active gels has been introduced \cite{LeGoff:2002p34076,Kruse:2004p10157,Joanny:2007p26677}, with the structure formation in the cytoskeleton being a prime example \cite{Lau:2003p34979,Joanny:2009p33206,Fletcher:2009p35180}. Active gel systems rely on pattern forming mechanisms very different from other soft materials: unlike structure formation in passive out-of-equilibrium systems dominated by nucleation and growth processes, the intriguing dynamical properties of active gels are the result of the interplay between active force generation and force dissipation in the (visco-)elastic environment \cite{Guerin:2010p32869}.

A minimal \emph{in vitro} system of a dynamic cytoskeletal network consists of the combination of only three components: actin filaments, myosin-II motor filaments and crosslinking proteins \cite{Kane:1983p5461,Janson:1991p18885,Liverpool:2009p5447}.
In a parameter regime where still homogenous networks are observed, the effect of molecular motors can directly be detected by the bulk properties of the networks \cite{Koenderink:2009p2973,Bendix:2008p5329}. In another concentration regime local and quasi static structures evolve \cite{Backouche:2006p537,Smith:2007p19085}. Yet, cells rely on highly heterogenous dynamic structures. It is this regime of active gels where cells are able to constantly reorganize their structure and mechanics to their local needs.

Here we present a minimal model system of an active gel consisting of actin filaments, fascin as crosslinking molecules and myosin-II filaments as a molecular motor. We identify and quantify  the key properties of the highly heterogenous and dynamic system and connect them to the underlying microscopic mechanisms. To this end we combine fluorescence microscopy, digital image analysis including recognition and tracking of actin structures and phenomenological simulations. At sufficiently high myosin-II concentrations, we observe a coarsening process, which results in a dynamic steady state, which is characterized by a broad cluster size distribution. Three distinct cluster sizes with their respective characteristic stability and dynamics can be identified. Only clusters of actin filaments above a critical size of 10--20\,\textmu{}m in diameter are able to grow and coarsen. Their growth up to several hundred \textmu{}m in diameter is limited by the concomitantly decreasing connectivity within the actin network and the subsequently decreased possibility of active transport. This dependence of the overall mobility on the growth and coarsening rate proves robust for all varied parameters, suggesting this to be a generic feature of such active networks. The transport and thus growth dynamics is anomalous, reminiscent of the intracellular dynamics in living cells.

\begin{figure}[h!]
 \centering
 \includegraphics[width=\textwidth]{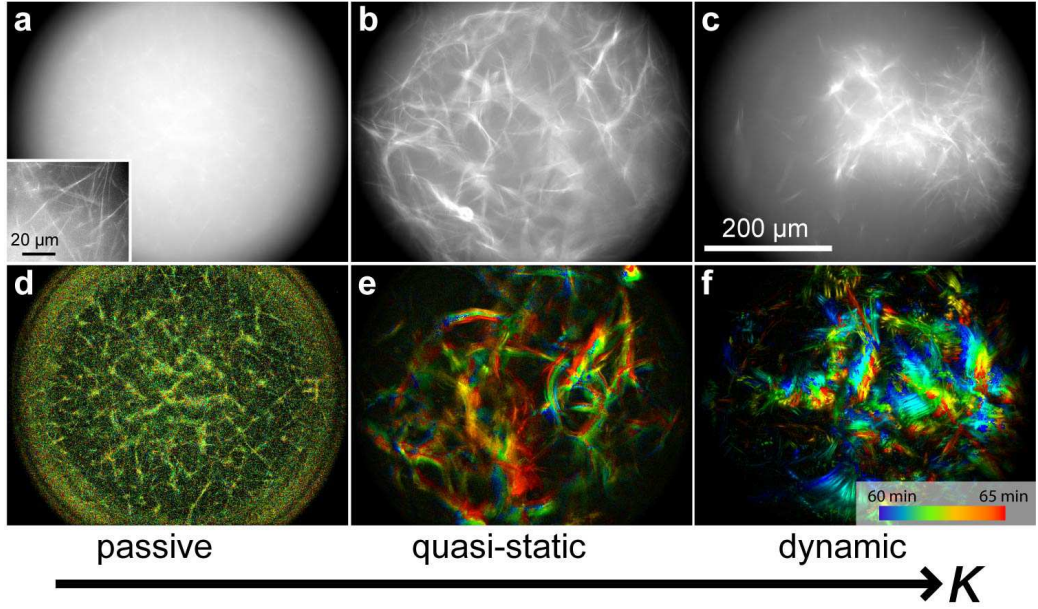}
 \caption{Structure of actin/fascin/myosin networks in dependence of $\kappa$. 
Fluorescence micrographs 90\,min after initiation of polymerization (\textbf{a--c}) and colored time overlays for 60--65\,min after initiation of polymerization (\textbf{d--f}) of 1\,\textmu{}M actin, 1\,\textmu{}M fascin at different myosin:fascin ratios are shown. (\textbf{a,d}) Without myosin, a passive actin/fascin network with thin, uniform bundles is observed. Only small fluctuations occur resulting in a homogenously colored timeoverlay. (\textbf{b,e}) Low myosin concentrations ($\kappa = 1:50$) exhibit quasi-static huge clusters spanning the whole field of view (regime i). The overall network structure remains constant, and only minor movements of the whole network are observable in the time overlay. (\textbf{c,f}) At high myosin concentrations, large, dense clusters and small clusters coexist in a dynamic steady state ($\kappa = 1:10$; regime ii). These dynamic reorganizations result in heterogeneous displacements in the time overlay. }
 \label{fig:phases}
\end{figure}

\paragraph*{Results}
\emph{In vitro}, actin filaments in the presence of the crosslinking molecule fascin at a molar ratio of $1:1$ assemble into a network of stiff and rigid bundles with a well defined bundle thickness of about 20 filaments per bundle and lengths of up to several hundred micrometer (Fig.~\ref{fig:phases}a) \cite{Claessens:2008p3149}. Thermal excited motions of the bundles are barely visible and the structure remains stable for hours although permanent unbinding and rebinding of the crosslinking molecules occur \cite{Lieleg:2008p771}.

Once the system is switched from a passive to an active state by the presence of myosin-II filaments and ATP, the network's structure and dynamics are drastically changed. Thereby, it is the concentration ratio between active components and static crosslinkers $\kappa=c_\mathrm{myosin}:c_\mathrm{fascin}$ that determines the network properties (Fig.~\ref{fig:phases}). The actin polymerization triggers a structure formation process that can last for up to 90 minutes resulting in a dynamic steady state.
This dynamic steady state shows a characteristic size distribution of the evolved patterns and distinct dynamical properties, both of which crucially depend on the choosen parameter set.
Roughly, two different steady state scenarios can be observed: i) a quasi-static regime and ii) a highly dynamic regime.
A prevalence of crosslinkers ($\kappa \lessapprox 1:50$) leads to the quasi-static network regime (i) with only minor reorganizations taking place (Fig.~\ref{fig:phases}b,e and supplemental movie~S1).
In this regime forces of the motor filaments are not sufficient to induce any large scale dynamics in this crosslinker dominated network. Consequently, a single percolated network with heterogenous bundle thicknesses is formed. Only small patches, which are stabilized by few crosslinkers can be displaced by the relatively small number of motor filaments. 

\begin{figure}[h!]
 \centering
   \includegraphics[width=\textwidth]{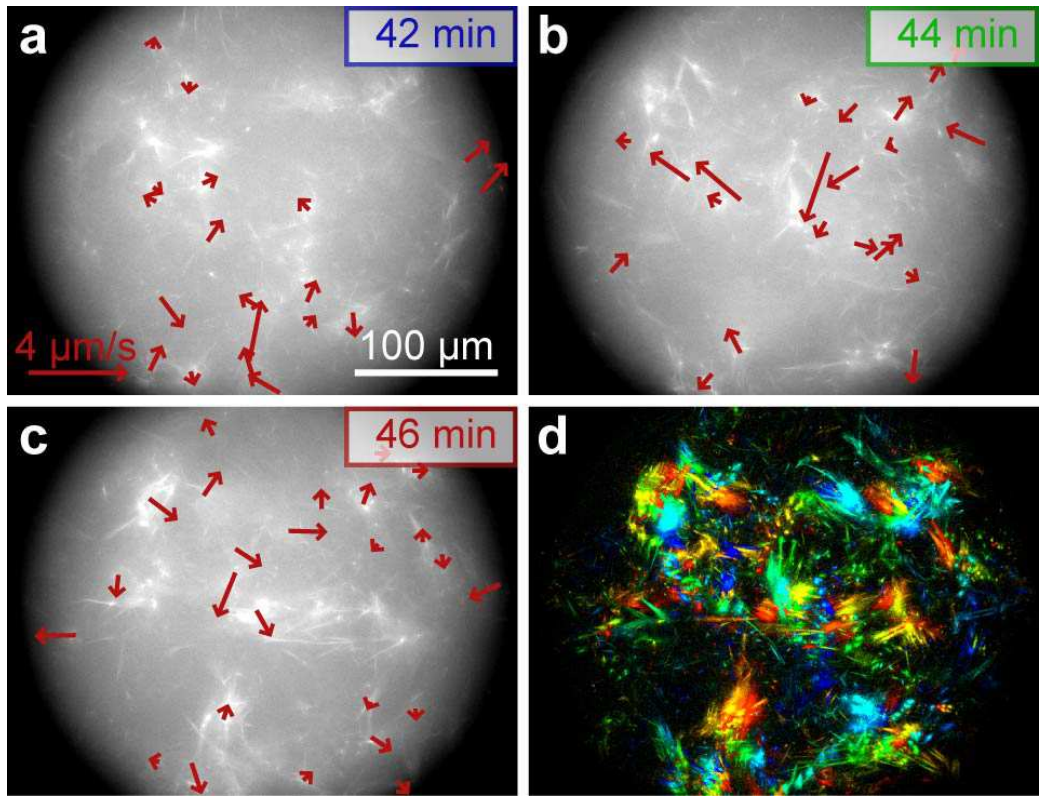}
    \caption{Dynamics of actin/fascin/myosin networks. 
      Fluorescence micrographs (\textbf{a--c}) and the resulting color time overlay (\textbf{d}) of active actin networks at indicated times after initiation of polymerization show a drastic structural rearrangement within the network (1\,\textmu{}M actin, 1\,\textmu{}Mfascin, 0.1\,\textmu{}M:myosin, 1\,mM ATP). Arrows indicate the cluster velocities. The overlay (\textbf{d}) represents 42--46\,min after polymerization in blue to red. }
 \label{fig:dynamics}
\end{figure}

Only in a parameter range where the influence of both components, passive and active crosslinkers is balanced ($\kappa \gtrapprox 1:50$), a highly dynamic regime (ii) of structure formation and constant reorganization can arise. Here the concentration of molecular motors is high enough to readily disrupt the network (Fig.~\ref{fig:phases}c). Instead of well defined bundles, condensed and interconnected actin-fascin clusters with diameters of up to hundreds of micrometer and variable shapes emerge (see methods). Once formed, the clusters are constantly subjected to the action of molecular motors (Fig.~\ref{fig:phases}f and supplemental Fig.~S1a--b). As a consequence, they are highly mobile and move in a succession of persistent runs with a mean velocity of 0.5\,\textmu{}m/s (Fig.~\ref{fig:dynamics} and supplemental movie~S2). The actin structures stabilized by the crosslinking proteins ensure a certain connectivity between individual clusters which is mandatory for a force percolation and the emergence of highly mobile and dynamic structures.

\begin{figure}[h!]
 \centering
 \includegraphics[width=.7\textwidth]{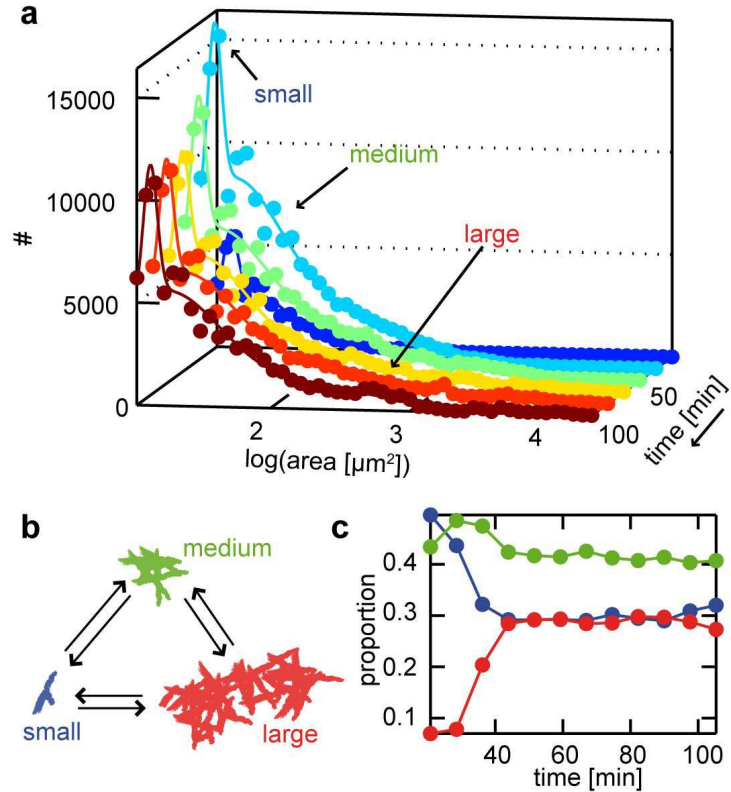}
 \caption{Time evolution of the cluster size distribution. 
(\textbf{a}) After polymerization, three types of clusters with distinct sizes emerge. The cluster sizes are determined by their respective areas. The distribution of the cluster areas are calculated for 18.4\,min time intervals and represents the integrated total number of clusters found in the field of view in this time interval. The positions of small, medium or large clusters, respectively, are equal for each time period. This time evolution is robust for all studied actin concentrations (see supplemental Fig.~S2). (\textbf{b}) Material exchange occurs between clusters of all sizes.  (\textbf{c}) With time, the population of the different cluster sizes changes as seen in their number proportion: Small, homogenously sized clusters (blue, with area smaller than 17.8\,\textmu{}m$^2$) first fuse forming medium sized clusters (green, area larger than 17.8\,\textmu{}m$^2$ and smaller than 63.1\,\textmu{}m$^2$). These medium sized clusters further grow forming large and stable clusters (red, area larger than 63.1\,\textmu{}m$^2$). A dynamic steady state evolves from medium sized clusters being constantly destroyed and reformed from few small clusters, while large clusters are stable. %
}
 \label{fig:size}
\end{figure}

\begin{figure}[h!]
 \centering
 \includegraphics[width=\textwidth]{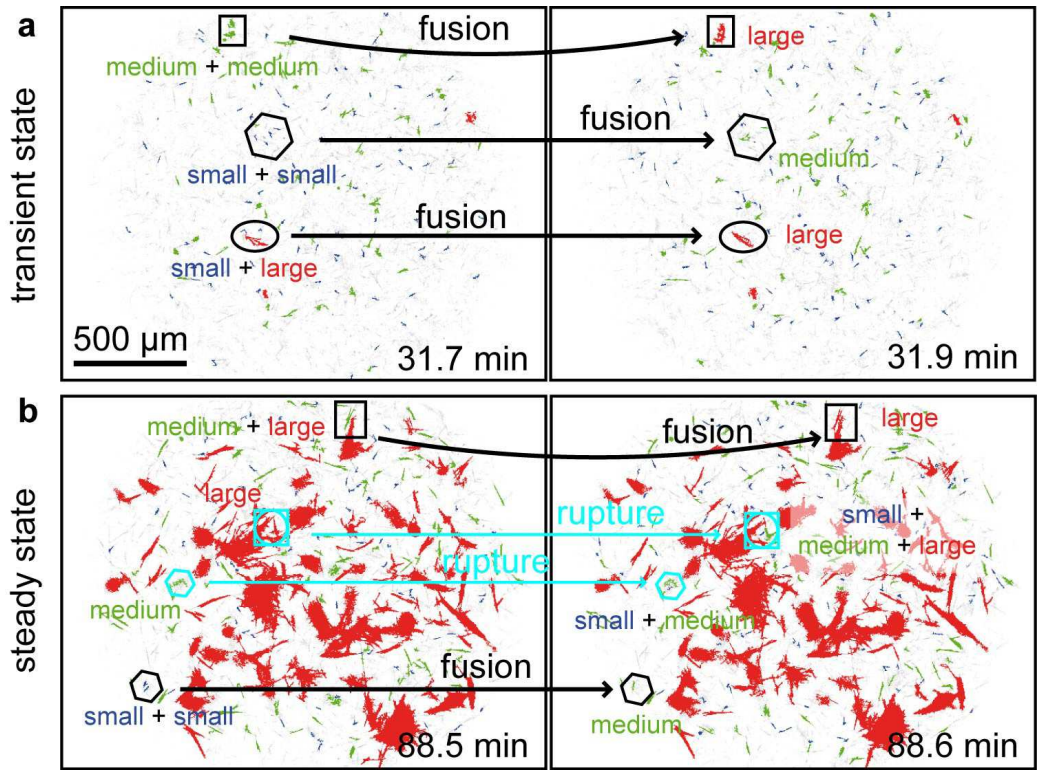}
 \caption{Dynamics of small, medium and large clusters. (\textbf{a}) In the initial transient state, predominately small (blue) and medium sized (green) clusters occur. They grow by fusion events: Small clusters coalesce to form medium sized clusters (black hexagon) or fuse with larger clusters (black circle). Large clusters are formed by fusion of medium sized clusters (black square). (\textbf{b}) In the dynamic steady state, fusion and rupture events are balanced. During rupture events, small or medium sized clusters are teared out of large clusters (cyan square and circle). Medium sized clusters can further be disintegrated into small clusters (cyan hexagon). See also  supplemental movie~S3.}
 \label{fig:cluster}
\end{figure}

In this regime, the cluster sizes in the steady state are broadly distributed (Fig.~\ref{fig:size}a, red). We define three characteristic sizes:  small, bundle like structures (average diameter $d\approx 6$\,\textmu{}m), medium sized clusters ($d\approx 10$\,\textmu{}m) and large clusters ($d\approx 34$\,\textmu{}m).
The time evolution of the cluster size distribution shows that initially predominantly small clusters consisting of several actin/fascin bundles are formed spontaneously (Fig.~\ref{fig:size} and \ref{fig:cluster}a, supplemental movie~S3). Once formed, these small clusters fuse to form medium sized clusters, yet disruption events of medium and large sized clusters result in their reformation (Fig.~\ref{fig:size}b). At the same time, small clusters are also stable as they are unlikely to be disrupted due to their small contact area. The resulting low probability of motor filaments binding in an appropriate manner to further disrupt the small clusters is limiting their disintegration. Consequently, a large number of small clusters can be found in the active system at all times.
Medium sized clusters are formed continuously by fusion of small clusters or disruption of large clusters. At the same time, they can either be annihilated by their adsorption to larger clusters or by disruption into small clusters. All these effects contribute to the fact that the total number of medium sized clusters predominates (Fig.~\ref{fig:size}c), although individual medium sized clusters are intrinsically unstable (Fig.~\ref{fig:cluster}b).
The large clusters are stable over time and increase their size by fusing events with clusters of all sizes. Small parts of them are torn out frequently without compromising their structural integrity.
Thus, over time a dynamic steady state of cluster size distributions evolves with a frequent exchange of material between the different cluster sizes (Fig.~\ref{fig:size}c and \ref{fig:cluster}b).

\begin{figure}[h!]
 \centering
 \includegraphics[width=.7\textwidth]{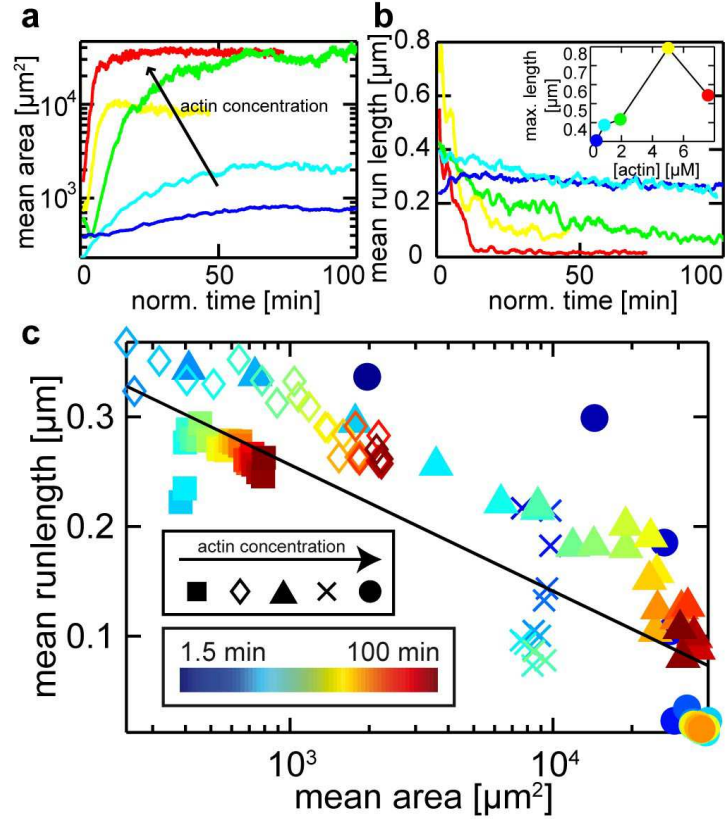}
 \caption{Dependence of the mean run length on the cluster size. 
(\textbf{a}) The growth of the mean cluster size in time depends on the actin concentration (1\,\textmu{}M fascin, $\kappa = 1:10$). Due to different polymerization times $t_0$, the time traces are normalized in time. (0.5\,\textmu{}M (blue) $t_0 = 33$min, 1\,\textmu{}M (cyan) $t_0 = 29$min, 2\,\textmu{}M (green) $t_0 = 23$min, 5\,\textmu{}M (yellow) $t_0 = 3$min, 7.5\,\textmu{}M (red) $t_0 = 1$min). (\textbf{b}) The mobility of the clusters represented by the mean run length per cluster decreases over time. This is more prominent the higher the actin concentration is (colors as in a). The initial maximal mobility of the system increases with the actin concentration (\textbf{Inset}). At the highest actin concentration studied, coarsening is almost instantaneous, that the maximal run length can only be determined for a later time point (red point).
(\textbf{c}) The mean run length decreases logarithmically (solid line to guide the eye) with the cluster size represented by the mean cluster area. This universal finding is found for different actin concentrations at 0.1\,\textmu{}M myosin-II ( 7.5\,\textmu{}M - circles, 5\,\textmu{}M - crosses, 2\,\textmu{}M - triangles, 1\,\textmu{}M - open diamonds, 0.5\,\textmu{}M - squares) or for different times after initiation of polymerization (blue to red corresponds to 1.5 to 100\,min). Each point represents a time average over 3\,min.
}
 \label{fig:size-run}
\end{figure}

The chracteristic temporal evolution of the cluster size distribution shown in Fig.~\ref{fig:size} is very robust: it is observed for all studied actin or myosin concentrations.
Only at the very low actin concentration of 0.5\,\textmu{}M large clusters hardly form and the mean clustersize barely increases (Fig.~\ref{fig:size-run}a and supplemental Fig.~S2a--b). At higher actin concentrations, large clusters grow to span the whole field of view decreasing the statistics of the cluster size distribution (supplemental Fig.~S2c--d). Furthermore, with increasing actin concentration (above 2\,\textmu{}M), the  formation of huge clusters results in a heterogenous network organization and only a part of the network structure can be observed in the field of view. This may lead to only fragmentary determination of the mean cluster sizes as observed for 5\,\textmu{}M actin in Fig.~\ref{fig:size-run}a. For all parameter variations, the system reaches a dynamic steady state, where the cluster size distribution and consequently the mean cluster size remain constant (Fig.~\ref{fig:size-run}a and Fig.~S1c).

The plateau value of the mean cluster size and the time to reach the plateau crucially depend on the actin concentration. Conceptually, a high actin concentration enhances the network connectivity and thus facilitates an effective self organization process. Subsequently, the steady state is reached faster as an increased actin concentration boosts the initial mobility in the system as measured by the mean run length per cluster (Fig.~\ref{fig:size-run}b, inset). Also, a rapid nucleation of small clusters is observable. For all actin concentrations, the subsequent formation of medium and large clusters results in a decline of the network's connectivity that in turn yields a decreased cluster mobility (Fig.~\ref{fig:size-run}b): The more material is accumulated in the large clusters, the less material is available to connect the different clusters with each other. Thus the tracks are lacking which are necessary for myosin-II to effectively transport actin structures, which determines the mobility of clusters. Importantly, at high actin concentrations higher average cluster sizes are observed  in the steady state (Fig.~\ref{fig:size-run}a). These clusters are less connected yielding a lower mobility (Fig.~\ref{fig:size-run}b).

Indeed, we observe a generic dependence of the mean run length on the cluster size and thus the connectivity of the network. Independent of the specific parameter set or the time the system has coarsened a logarithmic decrease of the mean run length per cluster with increasing cluster size (for details see methods) is found  (Fig.~\ref{fig:size-run}c). At the highest actin concentration used (7.5\,\textmu{}M), a macroscopic contraction into a single, non-mobile cluster is observed resulting in a freezing of the active network with only minor reorganizations taking place (Fig.~\ref{fig:size-run}c, circles and supplemental Fig.~S2c-d). However, at very low actin concentrations (0.5\,\textmu{}M), the cluster sizes remain predominantly small or medium sized. Consequently, the clusters remain in an active state even after 100 min. 

\begin{figure}[h!]
  \centering
  \includegraphics[width=\textwidth]{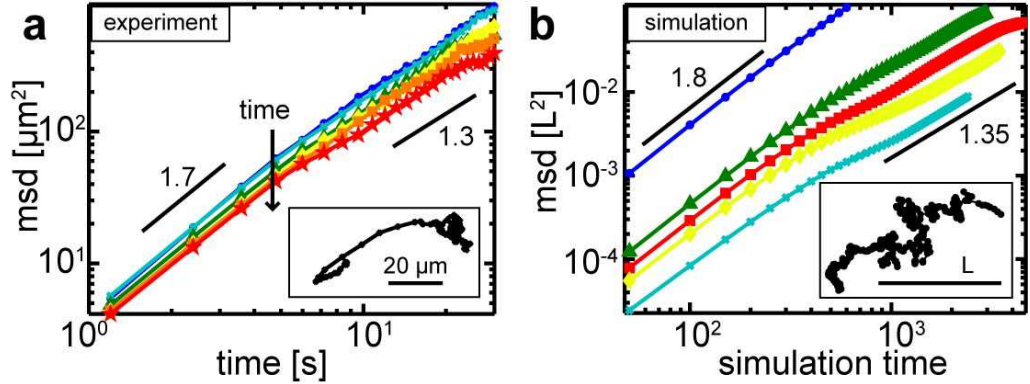}
  \caption{Mean square displacement (msd). (\textbf{a}) The ensemble averaged mean square displacement of the clusters at different time intervals shows superdiffusive behavior with a power law exponent larger than 1. Initially, in the transient state (17--35\,min, blue circles) a high superdiffusivity persists for long times. With increasing time (36--54\,min, cyan crosses; 55--72\,min; green triangles; 73--90, yellow diamonds; 91--109\,min, orange squares; 110--127\,min; red pentagrams), stalling events of the clusters for long times increase resulting in a biphasic behavior of the mean square displacement: a strong superdiffusivity due to local rearrangements at short times and only a slight superdiffusivity on larger time scales. Solid black lines are power laws with indicated exponents to guide the eyes. (\textbf{b}) Like in the experiment the transport in the simulations is characterized by a clear superdiffusive behavior. With time, individual clusters grow larger and move less, visble in the sample trajectory shown in the inset. This results in a successive decline in the mean square displacement exponent $\alpha$ and the experimentally observed biphasic behavior.}
  \label{fig:msd}
\end{figure}

The dynamics of the contraction is set by the transport mechanism within the network and is anomalous in nature. The dynamics of the individual structures in the network is best described by the mean square displacement (msd), which is computed for over 4000 individual clusters (for details see methods). For Brownian diffusion, the msd, $\langle |r(t+\tau) - r(t)|^2 \rangle \propto \tau^\alpha$, is expected to increase with time with $\alpha=1$ while $0 \le \alpha<1$ or $ 1<\alpha\le2$ are indicative of sub- or superdiffusion, respectively \cite{Metzler:2000p26002,Metzler:2000p41089}. For all actin concentrations studied, a clear superdiffusive behavior is found at all times (fig.~\ref{fig:msd}a and supplemental fig.~S3).  This superdiffusion with a powerlaw exponent larger than one, can be traced back to the complex alternation of runs and stalls in the individual trajectories (fig.~\ref{fig:msd}a, inset). While this dynamics with $\alpha \approx 1.7$ persists during the transient state where the structures evolve, a gradual decrease of the mobility is observed at longer times ($\alpha \approx 1.3$) with increasing clustering degree of the structures. At high actin concentrations, the long time behavior at late times becomes subdiffusive corresponding to immobile clusters showing only local rearrangements (fig.~S3b).

To relate the mesoscopic dynamics and the observed structure formation processes to the underlying microscopic mechanisms, the microscopic interactions are implemented in a numerical simulation: in a phenomenological approach, fascin bundles are modeled as monodisperse and polar rigid rods in a two dimensional geometry. The bundles are propelled and crosslinked, reflecting transport and binding processes. Both active and passive binding processes are subjected to rupture events based on a force dependent unbinding kinetics (for details see methods and supplemental online material).

\begin{figure}[h!]
 \centering
 \includegraphics[width=.9\textwidth]{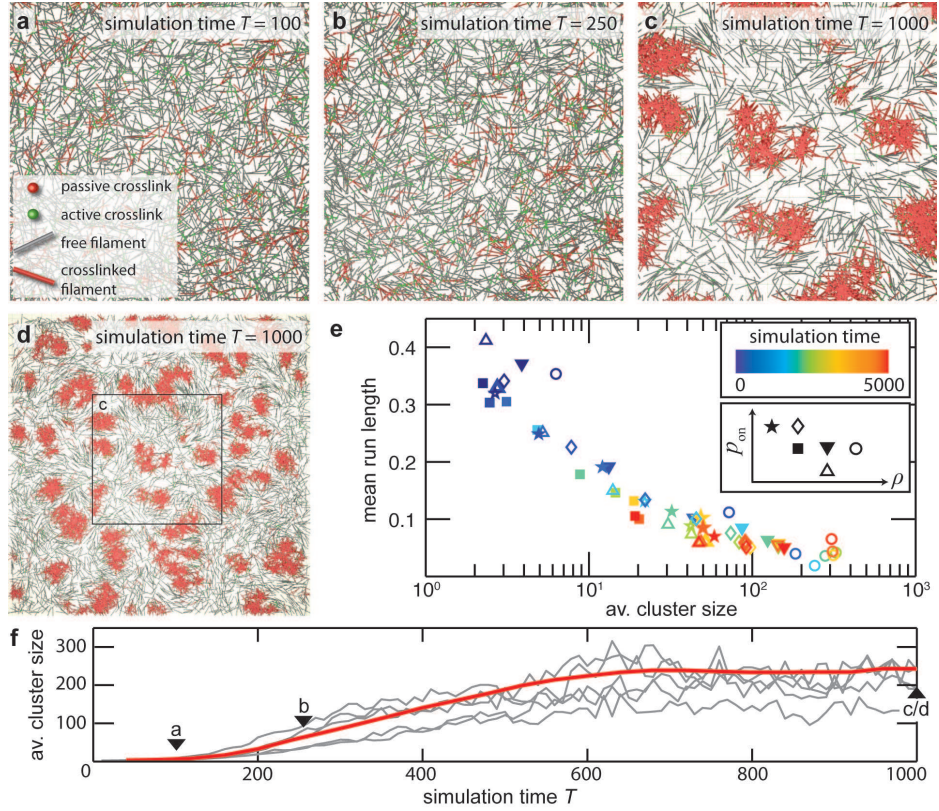}
 \caption{ Dynamics and coarsening behavior in the simulations. 
(\textbf{a--c}) show simulation snapshots for a density $\rho$ of 14 rods per $L^2$ (with $L$ being the rod length), a crosslinker unbinding rate $p_\mathrm{off} = 0.02$, a crosslinker binding rate $p_\mathrm{on} = 0.01$, a motor unbinding rate $r_\mathrm{off} = 0.01$, and a motor binding rate $r_\mathrm{on} = 0.1$ (for details see material and methods). The entire simulation box is shown in (\textbf{d}). In accordance with the experiment the simulation shows a rapid coarsenig process that leads to a steady state, where the average cluster size is stable in time (\textbf{f}). The more material is accumulated in the clusters, the lower is the connectivity and the motility in the system. This directly becomes manifest in the mean cluster run length which decreases with increasing cluster size (\textbf{e}). The mean run length is defined as the mean displacement of all clusters within simulation time intervals of $\Delta T = 10$. Like in the experiments the concomitant decrease of the mean run length with increasing cluster size is robust for a wide range of parameters (\textbf{e}). The actin density was varied from $\rho = 0.8$ -- $\rho = 1.6$ and the crosslinker on-rate from $p_\mathrm{on} = 0.007$ -- $p_\mathrm{on} = 0.013$. All lengths are measured in units of the rodlength $L=1$.}

 \label{fig:theo}
\end{figure}

By varying the ratio of motors and passive crosslinkers the experimentally observed phase behavior can be retrieved:
While an excess amount of passive crosslinkers leads to quasi-static structures with only minor reorganizations, dynamic structure formation processes only occur when motors and passive crosslinkers are relatively balanced.
Here, the interplay of active transport and crosslinking, leads to the formation of clusters of aggregated actin/fascin bundles.
Starting from a homogeneous initial state, the simulation shows a coarsening behavior in the course of which small structures like individual bundles or small clusters coalesce and gradually form larger structures (Fig.~\ref{fig:theo}a-d,f and movie~S4).
This is reflected in the temporal evolution of the cluster size distributions (Fig.~S4) that is in good aggreement with the experimantal findings.
Like in the experiment the initial cluster size distribution is characterized by a prevalence of small and medium sized clusters.
With time, in the course of the clustering process, the probability of finding large clusters increases.

The accumulation of material within the clusters during the observed coarsening process has profound consequences on the connectivity and hence on the mobility within the system. The more the coarsening has progressed, the smaller the connectivity becomes. Since a weak connectivity hinders the occurrence of persistent runs, large aggregated structures move less and exhibit smaller overall displacements. As a consequence, the mean run length in the simulation box declines with increasing cluster size in accordance with experimental findings (Fig.~\ref{fig:theo}e).

The dynamics of the clusters is anomalous and depends on the local connectivity and the cluster size, as large clusters that have grown over a long period of time and move less. This can be seen in analyzing the mean square displacement of the clusters at different times in the simulation (Fig.~\ref{fig:msd}b). The ensemble average over $1000$ clusters yields a clear superdiffusive behavior with an exponent between  $\alpha=1.8$ and $\alpha=1.35$, in excellent agreement with the experimental findings (Fig.~\ref{fig:msd}). The short time behavior is predominated by short runs, while the long time behavior at late times of the system gets dominated more and more by the dynamic arrest of the clusters. This explains the observed biphasic behavior of the msd in the steady state.

Like in the experiment, the stability of the clusters depends on their size. Large clusters are stable since they have grown sufficiently large and comprise a sufficiently large number of passive crosslinkers.
Motors continuously reorganize even large clusters internally by rupturing individual bonds. They are also able to tear out material from the clusters, yet they are not sufficiently strong and cooperative enough to entirely disintegrate these large structures.
Medium sized clusters on the contrary can readily be disintegrated by the action of molecular motors. Here, they find a large number of possible binding sites and thus can induce unbinding events of the relatively small number of passive crosslinks within the cluster.
For the smallest structures comprising only two to five individual bundles, motor-induced unbinding events only play a minor role: As small clusters do not offer enough binding sites for motors, the cluster disintegration predominantly occurs by stochastic unbinding events not involving motor proteins.

\paragraph*{Discussion}
Compared to other phase separation processes in soft materials like nucleation and growth of crystals \cite{Weitz:1984p29052}, phase separation in binary mixtures \cite{Langer:1971p53} or diffusion limited aggregation \cite{Viscek:1984p40817}, active gels in the parameter regime studied here show a remarkable dynamic behavior.

The incessant local input of mechanical energy at the smallest scales via myosin-II filaments drives a constant reorganization of the actin/fascin network through forced unbinding and rebinding events. In the course of this constant network reorganization, a highly dynamic steady state of aggregated clusters emerges whose characteristic properties like the mean cluster size and the activity crucially depend on the systems key parameters: the actin and motor filament concentration. The activity of the highly heterogenous system depends on its connectivity, as molecular motors need tracks on which forces can be exerted. This dependence of the activity on the connectivity of the networks opens up the possibility that already small perturbations can result in a large structural response by the identified self organization mechanisms of the active gel. Biochemical signaling as well as a mechanical stimulus would result in a shift in the phase diagram enabling a rapid, local and highly robust mechanosensing mechanism as observed in cytokinesis or upon mechanical stimulation \cite{Effler:2006p10475}.

Importantly, the transport dynamics recovered in the reconstituted system is in excellent agreement with the dynamics inside the cytoplasm of cells. Here too, a superdiffusive behavior is observable with exponents between 1.5 and 1.7 determined by different means \cite{Lau:2003p34979,Hoffman:2006p41353,Gallet:2009p41348}. While in cells the mechanism can only be linked to the energy consumption, presumably through the activity of molecular motors, the model system presented here allows the identification of the underlying microscopic mechanisms, which is the competition between binding and rupturing events evoked by crosslinking molecules and molecular motors.

While individual patterns like asters of cytoskeletal filaments have been successfully described both by mesoscopic models in the dilute regime \cite{Liverpool:2003p30078,Aranson:2005p27308} and macroscopic models \cite{Kruse:2004p10157,Juelicher:2007p26881}, the explicit modelling of the nonlinear dynamics of extended patterns and coarsening processes proves difficult. This is attributed to the inherent heterogeneity in many model systems which comprise a polymorphism of many coexisting patterns. Especially for generic approaches that are based on linear irreversible thermodynamics with geometric nonlinearities, it is therefore challenging to narrow down the parameter space and to decide which additional nonlinearities are important \cite{Juelicher:2007p26881}.

Other physiological cross-linkers, such as $\alpha$-actinin or filamin result already in the passive state in a much more complicated and even kinetically trapped network structure \cite{Schmoller:2009p28576}. It remains a formidable challenge to address the effect of active molecular motors on the local structure and dynamics in such networks. It is the excellent accessibility of the self organization principles and dynamics on all levels of description -- from the molecular mechanisms to large scale macroscopic pattern formation -- that makes the presented system based on fascin, myosin-II and actin to a versatile benchmark for the exploration of this broad material class of active gels. It may be the starting point for an \emph{in vitro} reconstitution of cellular dynamics by a bottom up approach.

\paragraph*{Methods}
\textbf{Protein purification} 
Myosin  \cite{Margossian:1982p24829} and G-actin \cite{Spudich:1971p24798,MacLeanFletcher:1980p24822} are extracted from rabbit skeletal muscle. Actin is fluorescently labelled with Alexa Fluor 555 succinimidylester (Invitrogen) with 25\,\% degree of labelling.
Recombinant human fascin is purified from \emph{E.coli} BL21-CodonPlus-RP and stored at -80\,$^{\circ}$C in 2\,mM Tris/HCl (pH~7.4), 150\,mM KCl at 64\,\textmu{}M \cite{Vignjevic:2003p24834}.

\textbf{Fluorescence imaging}
Actin in presence of indicated concentrations of myosin-II and fascin is polymerized by adding one-tenth of the sample volume of 100 \,mM imidazole, 2\,mM CaCl$_2$, 30\,mM MgCl$_2$. The ATP concentration is kept constant at 1\,mM ATP by adding 20\,mM creatine phosphate and 0.1\,mg/mL creatine phospho kinase (Sigma). 3\,mg/mL casein is added to prevent any surface interactions. In this assay buffer, myosin readily polymerizes into filaments with a mean length of 0.6\,\textmu{}m. Samples are enclosed to hermetically sealed chambers to eliminate any drift in the network.
All data are acquired on a Zeiss Axiovert 200 inverted microscope with either a 10\,x (NA 0.2) long distance objective or a 40\,x (NA 1.3) oil immersion objective. Images are captured at 0.84\,frames/s with a charge-coupled device camera (Orca ER, Hamamatsu) attached to the microscope via a 0.4\,x camera mount.

\textbf{Image processing}
Images are background subtracted in ImageJ. To identify individual clusters, an intensity threshold value is applied in ImageJ to generate a binary image. The threshold value is set such that the noise level remains constant for all studied samples. The cluster size is determined by their areas. Diameters are square roots of their respective area. 
The clusters are traced over time using the IDL tracking algorithm by John C.~Crocker \cite{Crocker:1996p44059} for the intensity weighted centroid cluster positions using Matlab R2008b (The MathWorks, Inc.). To minimize tracking artefacts, the trajectories are subjected to a gliding average over 4 frames. For each parameter set clusters were automatically identified and traced for 1-2 hours with a time resolution of 1.19\,s. As clusters are frequently merging or annihilated (as described in the text), or move out of or into the field of view up to 10,000 clusters per experiment were identified, while per frame not more than about 100 clusters are observed simultaneously.

To characterize the activity in the system, we calculate the mean run length per cluster. To this end, the trajectories are divided into "runs" and "stalls": Runs are identified as movements between two frames with velocities larger than 0.36\,\textmu{}m/s and a change in direction smaller than 30\,$^\circ$. The run length is defined as the length of a continuous run over at least 2 successive frames. The mean run length per cluster is calculated in 3\,min time intervals which is significantly longer than typical run times (10\,s average). 

The mean square displacement is calculated for individual traces as $\langle r^2(\tau) \rangle = \langle (r(t+\tau) - r(t))^2 \rangle$ and ensemble averaged subsequently for 18\,min time slots.

\textbf{Simulation}
In a minimal approach fascin bundles are modelled as monodisperse and polar rigid rods in a quasi two dimensional geometry.
Despite being two dimensional, no excluded volume effects are taken into account.
The bundles are actively propelled by motors and crosslinked via passive crosslinkers. Both active and passive bindings are subjected to forced unbinding events. The explicit dynamics of motor proteins and crosslinkers is not taken into acount; if two filaments overlap active or passive binding events occur based on probabilistic interaction rules. In a similar manner unbinding processes are calculated. The displacements that arise due to the action of molecular motors are calculated based on generic velocity models similar to Ref.~\cite{Liverpool:2003p30078}. For details, see supplementary online material.

\paragraph*{Acknowledgments}
We greatfully acknowledge technical support by M.~Rusp and G.~Chmel. 
We greatfully acknowledge the financial support of the DFG in the framework of the SFB 863, and partial support in the framework of the German Excellence Initiative by the 'Nanosystems Initiative Munich' and the 'Institute of Advanced Studies' (TUM-IAS). S.K and V.S. thank the 'International Graduate School for Science and Engineering'. V.S. acknowledges support from the Elite Network of Bavaria by the graduate programme CompInt. 

\paragraph*{Author contributions}
S.K  and A.R.B. designed experiments, performed and analyzed experiments. V.S., S.K. and A.R.B. conceived, performed and analyzed the simulations and wrote the paper.

\clearpage

\section*{Figures}

\end{document}